\pdfoutput=1
\documentclass[a4paper, 10pt, times, USenglish, 1p]{elsarticle}%
\biboptions{sort}%
\usepackage{babel}%
\usepackage{amsmath}%
\usepackage{amssymb}
\usepackage{subcaption}%
\usepackage[separate-uncertainty=true, per-mode=symbol]{siunitx}%
\usepackage{tensor}%
\usepackage[colorlinks]{hyperref}%
\usepackage[nameinlink]{cleveref}%
\renewcommand{\d}{\mathop{}\!\mathrm{d}}%
\newcommand{\op}[1]{\widehat{#1}}
\journal{Nuclear Physics A}%
\makeatletter%
\def\ps@pprintTitle{%
	\let\@oddhead\@empty%
	\let\@evenhead\@empty%
	\def\@oddfoot{\footnotesize© 2024. Licensed under a \href{http://creativecommons.org/licenses/by-nc-sa/4.0}{Creative Commons Attribution-Noncommercial-ShareAlike 4.0 International License}.}%
	\let\@evenfoot\@oddfoot%
}
\makeatother%
\begin{document}%
\begin{frontmatter}%
	\title{\texorpdfstring{Towards next-generation optical potentials\\for nuclear reactions and structure calculations}%
		{Towards next-generation optical potentials for nuclear reactions and structure calculations}}%
	\author{\texorpdfstring{Salvatore Simone Perrotta\corref{cor1}}{Salvatore Simone Perrotta}}%
	\ead{perrotta2 at llnl dot gov}%
	\author{Cole Davis Pruitt}%
	\author{Oliver C.~Gorton}%
	\author{Jutta E.~Escher}%
	\cortext[cor1]{Corresponding author}%
	\address{Lawrence Livermore National Laboratory, 94550 Livermore, California, USA}
	\date{October 19, 2024}
	\begin{abstract}%
        Optical-model potentials (OMPs) are critical ingredients for basic and applied nuclear physics. Present-day computational capabilities allow us to generate data-driven nucleon-nucleus OMPs that are non-local and exactly dispersive (as theoretically required to be), include statistically-sound uncertainty quantification, and are trained on both scattering and bound-state data from a wide area of the nuclear chart. Combined together, these features allow for significant improvement in fidelity and extrapolative power of the model.
		Here, we present preliminary work toward the development and training of such an OMP. %
        The capability of the model to describe data at this first stage is encouraging.
	\end{abstract}%
\end{frontmatter}%
\section{Introduction}\label{secIntro}
	
	Optical-model potentials (OMPs) are a model for the interaction between two nuclei. Their main purpose is to accurately approximate the interaction in the ``low-energy nuclear physics'' regime (collision energy $\lesssim \qty{200}{MeV}$) such that calculations remain mathematically and computationally affordable.
	Formally, %
	the OMP for a given pair of reaction partners is defined, within the Feshbach formalism \cite{Dickhoff2019,Moro2019}, as the projection of their full many-body interaction on the elastic channel, where the internal state of the nuclei is the same in the initial and final state of the full system. The OMP thus only acts on the coordinate of the motion between the nuclei's center-of-mass, reducing the equations of motion to a one-body quantum problem.
	Carrying out exactly the projection operation that generates the OMP is not viable in practice, %
    and approximations have to be sought. However, the formal treatment reveals several interesting features that the exact OMP must possess: %
	\begin{itemize}
	\item The potential has an imaginary part. In a scattering process, this allows the absorption of wave-function flux from the elastic channel. The OMP can thus describe inclusive cross sections for non-elastic scattering. %
	\item The potential operator is non-local (non-diagonal in spatial representation) and energy dependent, which are typical signs of an effective interaction or equation of motion. %

	\item In addition to scattering properties, a nucleon-nucleus OMP defined at all energies, above and below the Fermi surface, can supply the one-body density for nucleons in the nucleus %
	and, thus, observables such as nucleon numbers, single-particle levels,
	binding energies, %
	radii, skins, momentum distributions, etc.~\cite{Mahaux1991,Atkinson2019PhD}.
	These are a probe for the OMP both close to the Fermi surface and at deeply-bound energies.

	\item The causality principle requires each matrix element of
    the potential operator, $\op U$, to follow a ``dispersion'' (Kramers-Kronig) relation in energy \cite{Toll1956,Atkinson2019PhD},
	\begin{equation}\label{eqRelazioneKramersKronig}%
		\op U(E) = \op U_0 + \op U_D(E) ,\quad
		\mathrm{Re} \op U_D(E) = \frac{1}{\pi} \mathcal{P}\!\!\int_{\mathbb R} \frac{\mathrm{Im} \op U_D(\mathcal E)}{E - \mathcal E} \d\mathcal E \ ,
	\end{equation}
	where $\mathcal{P}\!\!\int$ denotes the Cauchy principal value integral, and $\op U_0$ is any real, energy-independent potential. %
	An OMP %
	that obeys \cref{eqRelazioneKramersKronig} is called ``dispersive OMP" or ``irreducible self-energy".
	Note how, in \cref{eqRelazioneKramersKronig}, $\mathrm{Re} \op U_D$ at any energy %
	is related to $\mathrm{Im} \op U_D$ at all energies, %
	revealing that scattering and bound-state properties of a system are not independent.
	\end{itemize}
	
	To practically determine an OMP, essentially two paths are viable. Microscopic approaches derive the potential combining, under suitable approximations, different kinds of information on nuclear structure and (typically) nucleon-nucleon interactions. Phenomenological models assume a reasonable empirical form for the potential that depends on several free parameters that are fit to data.
	Hybrid %
	approaches are also possible.
    Due to the associated mathematical and computational complexity, non-locality and dispersivity are often not enforced. %
    In general, %
    most phenomenological OMPs %
    are trained on, and give reasonable predictions for, only stable nuclei, because scattering data are scarce or absent for unstable ones.

	In this work, we aim to
    develop %
    a new phenomenological nucleon-nucleus OMP, %
    one which is fully non-local and dispersive.
    Our model can be trained on both scattering and bound-state data (including properties %
    known experimentally for many unstable nuclei). Together with the physical constraints that we enforce, this can improve the model's predictive power %
    even outside of stability.
	
\section{Preliminary results}\label{secResults}

    The %
    approach adopted here is quite similar to that of ref.~\cite{Pruitt2023}, with the following exceptions.
    To reduce the computational costs of the fitting process during the current developmental stage of the work,
	we currently apply no outlier rejection, and the unaccounted-for uncertainties (due to OMP defects and underestimation of experimental errors) are fixed based on prior expectations, %
    rather than fitted.
    The OMP's uncertainties are underestimated as a result. %
	To limit the size of the computation, %
    our model is currently trained on only few nuclei at a time. %
    The scattering %
    data employed %
    for fitting %
    are a subset of the %
    data corpora %
	in ref.~\cite{Pruitt2023},
	excluding data below \qty{15}{\MeV} %
	as the compound-nucleus contribution to the cross sections is currently neglected.
	Additionally, we train the model on some bound-state properties: %
	particle numbers, root-mean-square charge radii from ref.~\cite{Angeli2013}, and nuclear binding energies
	computed combining data from refs.~\cite{Wang2021AME} and \cite{NIST_ASD}. %

	The current form of our potential employs the traditional uniform-sphere Coulomb and Woods-Saxon-based volume and surface terms, complex and energy-dependent, and a real and energy-independent Woods-Saxon-based spin-orbit term. Volume and surface imaginary terms are asymmetric around the Fermi energy, which is different from most dispersive OMPs available in literature and was found to be important to fit reasonably on both scattering and bound-state properties at the same time. Non-locality is expressed using the Perey-Buck form from ref.~\cite{Perey1962}, which is however treated exactly with a full non-local solver (not resorting to any local approximation).

    \begin{figure}[tbp]%
		\begin{subfigure}[b]{.48\linewidth}%
			\centering
			\includegraphics[keepaspectratio = true, width=\linewidth]{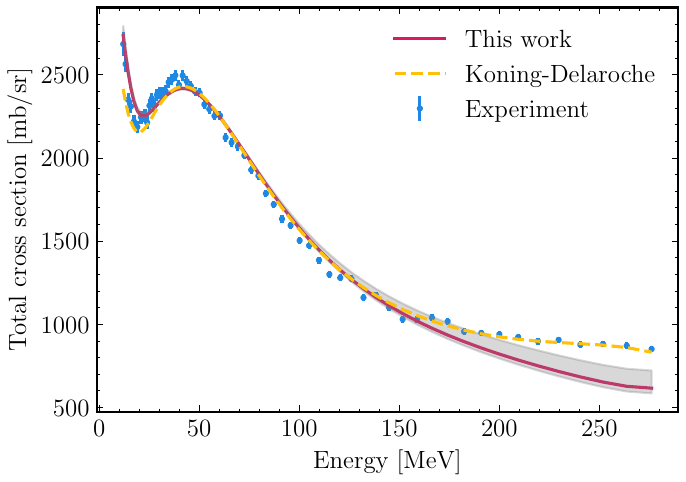}%
			\caption{\label{figResultsTCS48Ca}\nuclide[48]{Ca}+\nuclide{n} total cross-section (fitted between 15 and \qty{150}{MeV})}
		\end{subfigure}%
		\hfill
		\begin{subfigure}[b]{.48\linewidth}%
			\centering
			\includegraphics[keepaspectratio = true, width=\linewidth]{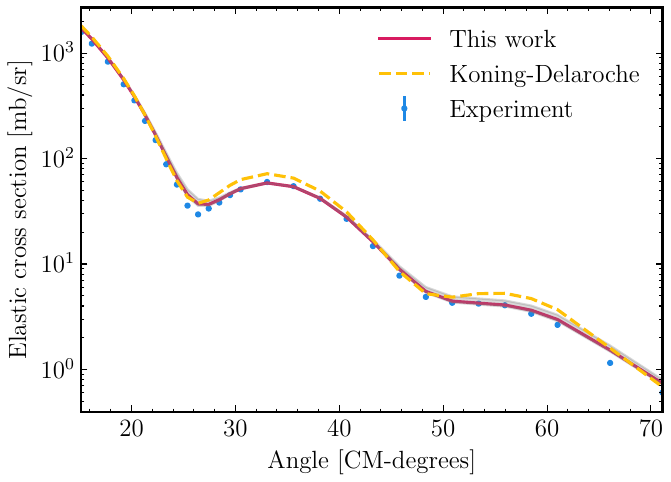}%
			\caption{\label{figResultsECS44Ca}\nuclide[44]{Ca}+\nuclide{p} elastic cross-section at \qty{65}{MeV} (predicted).}
		\end{subfigure}
		\vskip1.5ex
		\begin{subfigure}[b]{.48\linewidth}%
			\centering
	          \includegraphics[keepaspectratio = true, width=\linewidth]{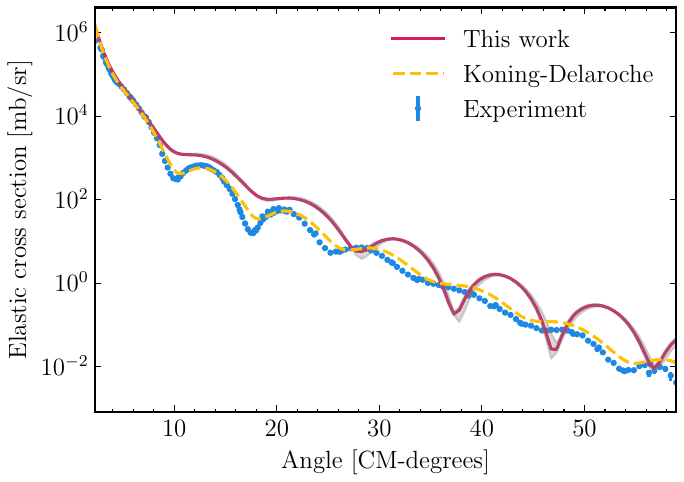}%
			\caption{\label{figResultsECS208Pb}\nuclide[208]{Pb}+\nuclide{p} elastic cross-section at \qty{200}{MeV} (predicted).}
		\end{subfigure}
        \hfill
        \begin{subfigure}[b]{.48\linewidth}%
			\centering
			\includegraphics[keepaspectratio = true, width=\linewidth, height=\textheight]{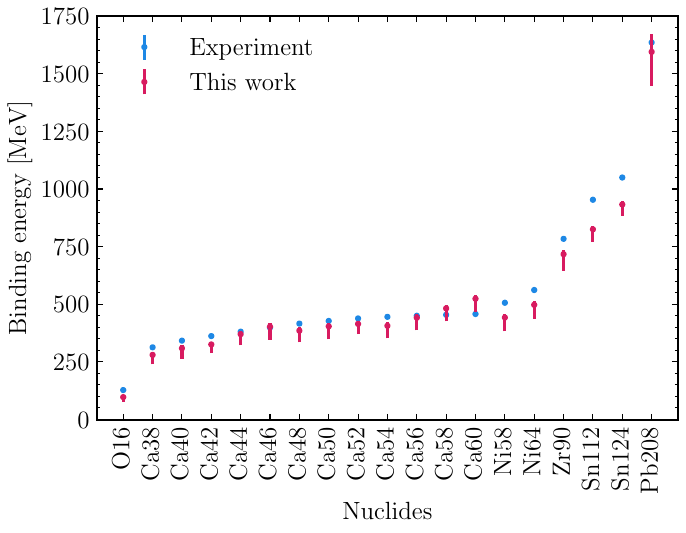}%
			\caption{\label{figResultsBE}Binding energy of several nuclides (fitted or predicted).}
		\end{subfigure}
		\caption{\label{figResults}%
			Representative preliminary results after training our OMP on non-polarized elastic scattering cross sections, total cross sections for neutrons and reaction cross sections for protons between 15 and \qty{150}{MeV} incident laboratory energy,
			as well as binding energies, %
			charge radii, charge and mass numbers, %
			for \nuclide[40,48]{Ca}, \nuclide[58,64]{Ni}, \nuclide[90]{Zr}, \nuclide[124]{Sn}, \nuclide[208]{Pb}.
			In each panel, blue points are experimental data from the sources mentioned in \cref{secResults}.
			Red points or solid lines are the fitted model's predictions, with corresponding $1\sigma$ uncertainties as error bars or grey bands. %
            Orange dashed lines are computed from the Koning-Delaroche OMP (which employs a different training dataset) %
            \cite{Koning2003}. %
            }
	\end{figure}
	\Cref{figResults} shows some preliminary results for a fit of our OMP on %
 7 nuclides across the chart. 
 Despite the limited training dataset, %
    the new OMP reaches a promising level of accuracy. %
 Performance for scattering data in the fitted energy range is comparable to the Koning-Delaroche global OMP \cite{Koning2003}, even for predictions on different stable nuclides outside our training dataset %
 (an example %
 is shown in \cref{figResultsECS44Ca}). %
 Extrapolation to %
 scattering energies beyond the fitted region is at present less accurate (see %
 \cref{figResultsTCS48Ca,figResultsECS208Pb}). For bound-state properties (which, currently, are under-represented in our fitting procedure), high-accuracy reproduction of data could not be reached: mean computed values for binding energies, shown in \cref{figResultsBE}, often deviate by about \qty{10}{\percent} from data.
	Notably, the model maintains a similar level of accuracy %
	even for predictions on different systems it was not trained to.
    In particular, the quality of the extrapolation of bound-state properties across the calcium chain (up to very neutron-rich isotopes) is remarkable.

	\section{Conclusions}
	Applications of the dispersive OMP typically only focus on one or few nuclides at a time, see for example refs.~\cite{Atkinson2019PhD,Pruitt2019PhD,Zhao2023}. A notable exception is the very recent work in ref.~\cite{Morillon2024}, presenting a global phenomenological dispersive OMP trained on neutron-scattering data. %
	We will attempt to go beyond that, by using a broader range of experimental data in our training and providing data-validated uncertainty quantification.
    To achieve the latter goal,
    we are working on implementing the more advanced uncertainty quantification tools from ref.~\cite{Pruitt2023} mentioned in \cref{secResults}. %
    This is paramount if one wants to meaningfully employ the model on observables for which experimental data is unavailable, as well as for sensitivity studies.
    Subsequently, we plan to obtain a global parametrization for our model, that is, one trained on and applicable to a wide area of the nuclear chart, which requires greater computational efficiency and/or capabilities.
    Thanks to its properties, it will be possible to train the OMP also on unstable nuclei, for which only basic information on nuclear structure is available.
    Our preliminary work towards this goal shows promising results.

\section*{Acknowledgements}
    We thank M.~C.~Atkinson for several helpful discussions on this project.
	This work was performed under the auspices of the U.S. Department of Energy by Lawrence Livermore National Laboratory under contract DE-AC52-07NA27344. %
\bibliography{bibliography}%
\end{document}